\documentclass[aps,prb,superscriptaddress,amsmath,amssymb,showpacs,twocolumn]{revtex4-1}

\usepackage{graphicx}
\usepackage[utf8]{inputenc}
\usepackage[T1]{fontenc}

\usepackage[english]{babel}
\newcommand{\beq}{\begin{equation}}
\newcommand{\eeq}{\end{equation}}
\newcommand{\beqa}{\begin{eqnarray}}
\newcommand{\eeqa}{\end{eqnarray}}

\newcommand{\bsub}{\begin{subequations}}
\newcommand{\esub}{\end{subequations}}

\newcommand{\rem}[1]{}
\newcommand{\refe}[1]{Eq.~(\ref{#1})}
\newcommand{\reff}[1]{Fig.~\ref{#1}}

\newcommand{\Imx}{I_{\rm mx}}

\begin{document}
\title{Sensitivity of the mixing current technique to detect nano-mechanical motion}
\author{Yue Wang}
\author{Fabio Pistolesi}
\affiliation{
Univ. Bordeaux, LOMA, UMR 5798,  Talence, France.\\
CNRS, LOMA, UMR 5798, F-33400 Talence, France.\\
}

\begin{abstract}
Detection of nano-mechanical displacement by transport techniques has reached high level 
of sensitivity and versatility. 
In order to detect the amplitude of oscillation of nano-mechanical 
oscillator a widely used technique consists to couple this motion capacitively to a single-electron transistor
and to detect the high-frequency modulation of the current through the non-linear mixing with an
electric signal at a slighltly detuned frequency. 
The method known as current-mixing technique 
is employed in particular for the detection of suspended carbon nanotubes. In this paper we 
study theoretically the limiting conditions on the sensitivity of this method. 
The sensitivity is 
increased by increasing the response function to the signal, but also by reducing the noise.
For these reasons we study systematically the response function, the effect of 
current- and displacement-fluctuations, and finally the case where the tunnelling rate of the electrons 
are of the same order or larger of the resonating frequency. 
We find thus upper bounds to the sensitivity of the detection technique. 
\end{abstract}
\date{\today}
\pacs{85.85.+j,73.23.-b,73.23.Hk}
%73.23.Hk 	Coulomb blockade; single-electron tunneling
%
\maketitle
%
%%%%%%%%%%%%%%%%%%%%%%%%%%%%%%%%%%%%%%%%
\section{Introduction}
%%%%%%%%%%%%%%%%%%%%%%%%%%%%%%%%%%%%%%%%
%
Nano-electromechanical systems have great potentials as ultra-sensitive
detectors for several physical quantities. 
Recent advances allowed to reach record sensitivity in mass 
sensing.\cite{ekinci_ultrasensitive_2002,lassagne_ultrasensitive_2008,chaste_nanomechanical_2012}
This has been possible by the detection of the frequency shift of ultralight oscillators when an additional 
mass is attached to it. 
Other exemples concern the detection of the tiny magnetic field generated by nuclear spins.
This can be done by the opto-mechanical detection of the force generated by the magnetic
dipoles,\cite{mamin_nuclear_2007} but also with electro-mechanical means,\cite{poggio_off-board_2008} or by coupling to two-level systes.\cite{rabl_strong_2009,arcizet_single_2011,puller_single_2013}
The force sensitivity of the device is then the limiting factor for the sensitivity, and again recent advances 
showed that it is possible to obtain record force sensing with carbon-nanotube oscillators.\cite{moser_nanotube_2014,moser_ultrasensitive_2013}
At the same time nano-mechanical oscillators can be so small that interaction between electronic and mechanical 
degrees of freedom may lead to new and unexpected 
phenomena\cite{blanter_single-electron_2004,blanter_erratum:_2005,armour_classical_2004,doiron_electrical_2006,pistolesi_tunable_2015} 
like the blockade of the 
current\cite{mozyrsky_intermittent_2006,koch_franck-condon_2005, pistolesi_current_2007,pistolesi_self-consistent_2008}, cooling\cite{pistolesi_cooling_2009,zippilli_cooling_2009,stadler_ground-state_2014} or 
unusual mechanical response.\cite{micchi_mechanical_2015,micchi_electromechanical_2016} 

In order to exploit nanomechanical resonators, or to study their properties, detection of mechanical motion is 
crucial. 
Most detection methods exploiting electronic transport are based on the high sensitivity of single-electron 
transistors (SET) to a variation of the gate charge.
By coupling capacitively the oscillator to the gate of the SET it is possible to detect the motion of the oscillator 
with a high accuracy.\cite{blencowe_sensitivity_2000}
The method has been used also to cool the oscillator by the back-action of the electronic transport.\cite{naik_cooling_2006}
The main difficulty of the method stems from the high frequency character of the oscillator motion that is 
typically in the 100 MHz-1 GHz range. 
Due to the high impedance of the SET,  it is more convenient to down-convert the signal to lower frequency
before extracting it. 
This can be achieved by non-linear mixing the mechanically generated modulation with a second high-frequency signal
injected between source and drain. 
The signal at the difference of the two frequencies can be extracted and measured. 
To our knowledge, for nanomechanical resonators this method was implemented in 
metallic SET by the group of A. Cleland back in 2003.\cite{knobel_nanometre-scale_2003}
It was later adapted to the detection of carbon nanotube by the group of P.L. McEuen.\cite{sazonova_tunable_2004}
It then became the method of choice for carbon nanotubes, leading to several breakthroughs: 
the observation of the first single-electron backaction effects in carbon 
nanotubes,\cite{lassagne_coupling_2009,steele_strong_2009} 
ultrasensitive mass detection,\cite{lassagne_ultrasensitive_2008,chaste_nanomechanical_2012}
the detection of the charge response function in quantum dot,\cite{meerwaldt_probing_2012} 
the detection of magnetic molecules\cite{ganzhorn_dynamics_2012,ganzhorn_carbon_2013}
and the observation of decoherence of mechanical motion.\cite{schneider_observation_2014}
The same method can also be implemented by frequency modulation.\cite{gouttenoire_digital_2010}
It is clear that the technique is powerful and that it will continue to be used
both for fundamental research and for applications. 
The question we want to address in this paper is which is the ultimate resolution that can 
be reached with this kind of detection.
In order to do this we investigated three main issues.
The first one is how to optimize the response function, that is the 
quantity $\partial \Imx/\partial x_m$, where $\Imx$ is the measured 
signal, the mixing current, and $x_m$ the amplitude of the mechanical oscillation.
The second one is to study the effect of current and mechanical fluctuations. 
These contribute to the fluctuation of the measured signal and in the end are at the 
origin of the signal to noise ratio.
The third is to consider the case of a mechanical oscillator with a resonating 
frequency $\omega_m$ faster than the typical tunneling rate of the electrons $\Gamma$. 
We will develop a theory of transport to obtain the mixing current for any 
ratio $\omega_m/\Gamma$.
The case of a metallic and single-electronic level SET will be considered in details
and explicit expressions will be given.

The paper is structured as follows: 
Section \ref{sec2} gives an introduction to the mixing technique 
and provides the expression of the detector gain $\lambda$.
Section \ref{sec3} analyze current and mechanical fluctuations giving 
general expressions.
Section  \ref{sec4} provides a general theory for the detector gain 
when the resonator frequency is comparable or larger than the typical tunnelling time.
Section \ref{sec5} and \ref{sec6} gives the explicit expressions for the 
response function and for the noise in the case of a metallic and a 
single electron SET.
Finally Section \ref{sec6} gives our conclusions.

%%%%%%%%%%%%%%%%%%%%%%%%%%%%%%%%%%%%%%%%

\section{Mixing technique and response function}
\label{sec2}

%%%%%%%%%%%%%%%%%%%%%%%%%%%%%%%%%%%%%%%%

\begin{figure}[tbp] % Requires \usepackage{graphicx}
\includegraphics[width=.95\linewidth]{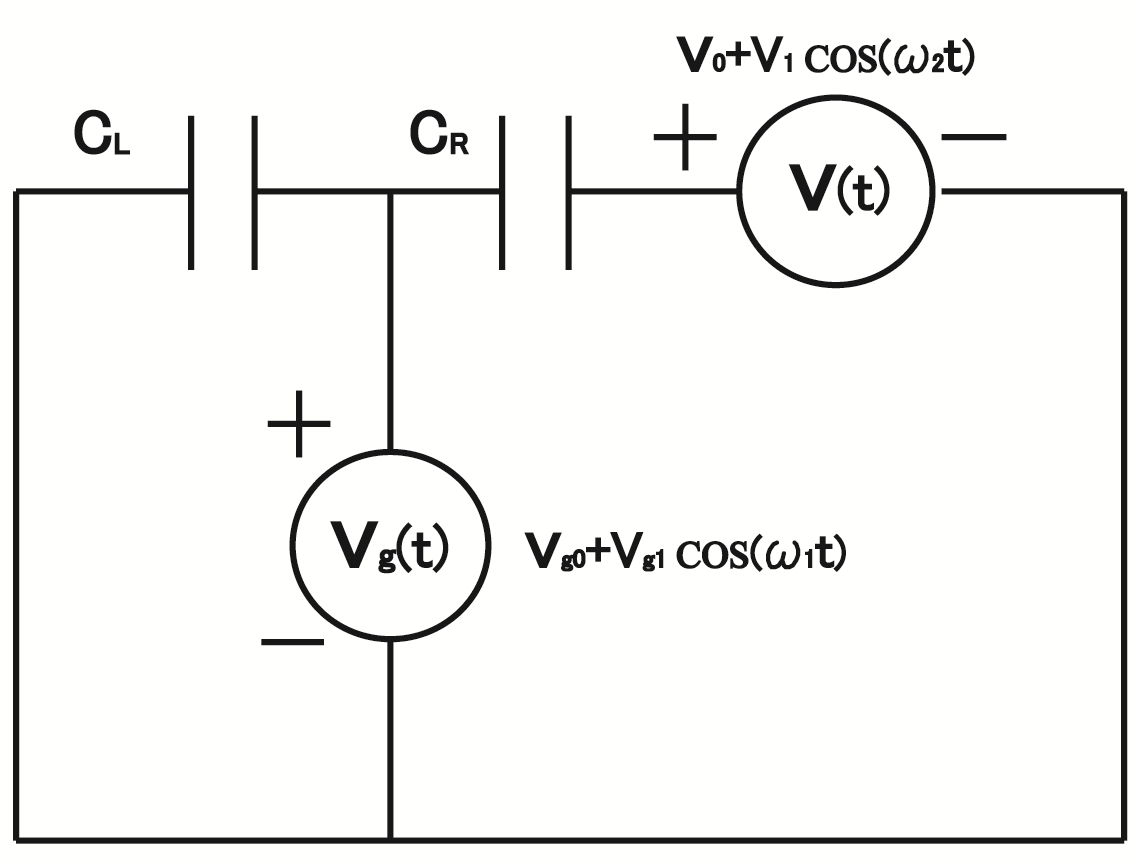}
\caption{Schematic of the typical experimental set-up used to 
measure the displacement of a mechanical oscillator by detection 
of the mixing current (adapted from Ref.~\onlinecite{sazonova_tunable_2004})
}
\label{schematic}
\end{figure}

Let us begin by describing the typical system used to measure 
the oscillation amplitude of a mechanical oscillator by detection of 
the mixing current.\cite{knobel_nanometre-scale_2003,sazonova_tunable_2004,lassagne_coupling_2009,steele_strong_2009}
As shown in \reff{schematic} a conducting oscillator is capacitively
coupled to the central island of a single-electron transistor: 
its displacement modulates thus the 
the gate capacitance $C_g(x)$, where $x(t)$ is the displacement 
of the oscillator.
%
%
%
%%%%%%%%%%%%%%%%%%%%%%%%%%%
%                                                                                %
%                              FIGURE 1                                 %
%                                                                                %
%%%%%%%%%%%%%%%%%%%%%%%%%%%
%
%
%%%%%%%%%%%%%%%%%%%%%%%%%%%%
%
%
We assume the presence of a single mechanical mode whose displacement 
is parametrized by $x$, a generalized coordinate with the dimensions of a length. 
We will consider that the SET is operated in the incoherent transport regime
valid for $\hbar \Gamma \ll k_B T$, where $\Gamma$ is the electron tunnelling rate
and $T$ the temperauture ($\hbar$ and $k_B$ are the reduced Planck constant 
and the Boltzmann constant, respectively).
This is the standard case for nano-mechanical devices.
The current $I$ through the device can be obtained by using the Master 
equation and in general it can be expressed as a function of the
source-drain bias voltage $V$ and on the gate charge 
$n_g=C_g(x) V_g/e$, where $V_g$ is the gate voltage
(see Appendix \ref{appA} for a short derivation).
The current reads thus:
\beq
	I=I(V, n_g) \,.
	\label{current}
\eeq
In this section we want to obtain the current response of the system when 
both $V$ and $V_g$ are modulated at two slightly different frequencies
$\omega_1$ and $\omega_2$, both much smaller than $\Gamma$.
We write
\beq
	V_g(t)=V_{g0}+V_{g1}(t) 
			\,,
	\qquad
	V(t)=V_0+V_1(t)  
	\,,
\eeq
where $V_{g1}(t)=V_{g1} \cos(\omega_1 t)$ and 
$V_1(t)=V_1 \cos(\omega_2 t)$.
Choosing $\omega_1$ close to the mechanical resonating frequency 
$\omega_m$ allows to drive the resonator, 
since the modulation of the gate voltage 
modulates the charge on the suspended part and thus induces an 
oscillating force [see also \refe{force} in the following].
For small driving amplitude the oscillator responds linearly to the external drive:
\beq
	x(t)=x_m \cos(\omega_1 t+\phi) ,
\eeq
where we always measure $x$ from its equilibrium position. 
(Note that in general $x_m$ and $\phi$ depend on the driving 
frequency $\omega_1$.)
The modulation of $V_g$ induces thus the following modulation of $n_g$ at linear order 
in the driving:
\beq
	n_g(t)=n_{g0} + {C_g V_{g1}\over e} \cos(\omega_1 t)+{C_g' x_m V_{g0 }\over e} \cos(\omega_1 t+\phi),
\eeq
where $C_g' \equiv dC_g/dx$. It is convenient to introduce a length scale by 
defining $L=C_g/C_g'$. 
From geometric considerations $L$ has to be of the order of the distance of the gate from the 
oscillator, thus typically undreds of nm.
The fluctuting part of $n_g$ can then be written as
\beq
	n_{g1}(t)= n_{g0} \left[ {V_{g1}\over V_{g0}} \cos(\omega_1 t)+{x_m\over L} \cos(\omega_1 t+\phi)\right]
	\,,
\eeq
where $n_{g0}={C_g V_{g0}/e}$.
The mechanical term $(x_m/L)$ has a strong frequency dependence close to the 
mechanical resonance, and can thus be distinguished by the back-ground electrostatic term 
$ (V_{g1}/V_{g0}) $.
The two contributions to the modulation of the gate charge can be combined in a
single cosine term: 
\beq
	n_{g1}(t)=n_{g1} \cos(\omega_1 t +\varphi)
	\,.
\eeq
Assuming now that the oscillator frequency $\omega_m$,
and thus also $\omega_1$ and $\omega_2$,
are much smaller than the typical tunneling rate $\Gamma$, 
one can use \refe{current} to obtain the time dependent 
current in presence of time-dependent $V$ and $n_g$.
For small modulation amplitude we  
Taylor expand \refe{current} to second order in $V_1$ and $n_{g1}$ obtaining 
\beqa
\lefteqn{	I(t)=
	I(V_0,V_{g0})+
	{\partial I\over \partial V} V_1(t)+
	{\partial I\over \partial n_g} n_{g1}(t)
	}
	\nonumber \\
	&&+
	{1 \over 2} {\partial^2 I \over \partial V^2} V_1^2(t)+
	{\partial^2 I \over \partial V \partial n_g} V_1(t) n_{g1}(t)+
	{1 \over 2} {\partial^2 I \over \partial n_g} n_{g1}^2(t)
	+\dots
	\nonumber \\	
	\label{expansion}
\eeqa
Only the term proportional to ${\partial^2 I/\partial V \partial n_g}$ has a component that 
oscillates at the frequency $\omega_\Delta=\omega_1-\omega_2$. 
This signal can be extracted by a standard lock-in technique that essentially allows to 
measure the quantity $\Imx$:
\beq
	\Imx^c=\int_0^{T_m} {dt\over T_m} I(t) \cos[\omega_{\Delta} t]
		\,.
\eeq
The other quadrature $\Imx^s$ with $\sin (\omega_{\Delta}t)$ is defined in a similar way.
Averaging over a long measurement time $T_m\gg 1/\omega_{\Delta}$ one obtains:
\beqa
	\Imx^c &=& {V_1 \over 4} {\partial^2 I \over \partial V \partial n_g}  \left[C_g V_{g1}+C'_g V_{g0} x_m \cos \phi \right] 
	\,,
	\label{MixCos}
	\\
	\Imx^s &=& -{V_1\over 4} {\partial^2 I \over \partial V \partial n_g}  C'_g V_{g0} x_m \sin \phi
	\,.
	\label{MixSin}
\eeqa
The detector gain with respect to the two quadrature of $x_m$ is thus:
\beq
	\lambda ={1 \over 4 e} {\partial^2 I \over \partial V \partial n_g}  C'_g V_{g0} V_1 \, .
	\label{response}
\eeq
It measures the sensitivity of the mixing current signal with respect to the two quadratures of $x$. 
This quantity depends on the particular bias conditions of the SET, and will be studied 
in some details in Section \ref{sec5} and \ref{sec6} for two explicit models.
Note also that in order to obtain $\lambda$ we need only the static expression for the 
current. This assumes that the electronic mechanism is much faster than the 
time dependence of the driving. In order to describe the case of a fast 
oscillator (to be discussed in Section \ref{sec4} we will need a detailed description of the charge dynamics,
and the response function will be no more expressed only in terms of 
derivatives of the static non-linear current voltage characteristics.

%%%%%%%%%%%%%%%%%%%%%%%%%%%%%%%%%%%%%%%%
 
\section{Effect of current and displacement fluctuations}
\label{sec3}

%%%%%%%%%%%%%%%%%%%%%%%%%%%%%%%%%%%%%%%%

Expression (\ref{response}) assumes a deterministic evolution of both the current and the 
displacement of the oscillator $x(t)$. 
In practice both quantity fluctuate, the first due to shot or thermal noise, and the second due to 
stochastic fluctuations induced either by the bias voltage or by the thermal fluctuations.
In general one can then write the value of $\Imx$ in a specific time region as follows:
\beq
	({\Imx^c})_n=\int_{n T_m}^{(n+1)T_m} [I(t)+\delta I(t) ]\cos(\omega_{\Delta} t) dt
		\,,
\eeq
(we write the expression for $\Imx^c$, the one for $\Imx^s$ is similar)
where $\delta I(t)$ and $I(t)$ are the stochastic and deterministic (in phase with the 
external drive) part, respectively.
We can define the time dependent mixing current as $\Imx^c(t)=({\Imx^c})_{[t/T_m]}$, where 
$[\alpha]$ stands here for the integer part of $\alpha$.
In terms of that the spectral density of the fluctuation of $\Imx^c$ reads: 
\beq
	S_{\rm mx}(\omega)=\int_{-\infty}^{+\infty} dt e^{i \omega t} 
		\left[ \langle \Imx^c(t) \Imx^c(0) \rangle - \langle \Imx^c\rangle^2\right]
		\,.
\eeq
We assume that the measuring time is much longer than any correlation time of the 
quantity $\delta I (t)$.
Different sections of the measurement time are thus uncorrelated and we can write:
\beqa
	S_{\rm mx}(\omega)&=& \int_0^{T_m} dt e^{i \omega t}  
	\int_0^{T_m} {dt_1\over T_m} 
	\int_0^{T_m} {dt_2\over T_m}
	\nonumber \\ 
	&&
	\cos(\omega_{\Delta}  t_1)\cos(\omega_{\Delta} t_2)
	\langle \delta I(t_1)\delta I(t_2) \rangle
		\,.
\eeqa
Defining $S_{I}(\omega=0)= 2 \int_{_\infty}^{+\infty} dt  \langle \delta I(t)\delta I(0) \rangle$  
(the numerical factor 2 is conventional for the current-noise spectrum) we have
\beq
	S_{\rm mx}(\omega)=
	{1\over 4} S_{I}(\omega=0){(e^{i\omega T_m}-1) \over i\omega T_m} \approx 
	 {1\over 4} S_{I}(\omega=0)
\,.
\eeq
Thus the mixing-current low-frequency noise is given simply by the low-frequency current noise
spectrum $S_{II}$.
The factor of 4 comes from a different definition of the correlation functions and from the fact that 
we are collecting a single quadrature.
The current noise can have different sources, we consider in the following the two main ones.

\subsection{Shot-noise  and thermal current fluctuations}

The current fluctuates due to the discrete nature of the the charge.
This is characterized by the current-spectral function (for time-independent bias and gate voltages):
\beq
	S^{\rm shot}_{I}(\omega)= 2\int dt e^{i \omega t} \langle \delta I(t) \delta I(0) \rangle
		\,,
\eeq
where $\delta I(t)= I(t)-\langle I \rangle$.
For the case of a SET the current spectral function is well known.\cite{korotkov_intrinsic_1994}
As shown there it has a frequency dependent part at low frequency 
on the scale of the typical tunneling rate $\Gamma$.
This implies that the correlation function is short ranged with respect to the measuring time 
$T_m$. 
Actually it is typically even short ranged with respect to the time dependence of 
$x$ and of the $V$ or $V_g$ potentials. 
Its value can thus be obtained adiabatically, by assuming these parameters to be static.
We only need its low frequency part that can, in general,  be expressed 
in terms of the Fano factor $\cal F$ and the current $I$:
\beq
	S^{\rm shot}_{I}(\omega=0) = 2 {\cal F} I e
\eeq
where ${\cal F}$ depends on the details of the SET. In the tunnelling limit of 
uncorrelated tunneling  ${\cal F}=1$, in most other cases the Fano factor is typically
of the order of 1. 

\subsection{Displacement fluctuations}

The electrons that cross the structure modify the charge on the gate that in turn modifies the 
force acting on the oscillator. 
This stochastic force, that has the same origin of the current-shot noise, induces fluctuations 
of the displacement, that changes in a much slower way, since the oscillator responds 
to an external force on the time scale given by its damping coefficient $\gamma$.\cite{armour_classical_2004,armour_current_2004,usmani_strong_2007,pistolesi_current_2007}
In order to keep the assumption that different averages over the measuring times are uncorrelated
one needs $T_m\gamma \gg 1$.
In principle, for very high-$Q$ resonators the approximation should be reconsidered. 

Let's begin by considering the force acting on the oscillator as a consequence of a variation of the 
charge on the gate. 
A recall of the basic expressions for the electrostatic energy is given in the Appendix \ref{appA} 
and Fig.3  there shows the electrical scheme. 
The force acting on the oscillator is given by the derivative of the electrostatic energy 
performed at constant charge: 
\beq
	F=-Q_g^2 {\partial \over \partial_x} {1 \over 2 C_g(x)}={Q^2_g C'_g \over 2 C^2_g},
	\label{force}
\eeq
where $Q_g$ is the charge on the gate voltage (Fig.3).
The fluctuation of the force  $\delta F(t)$ due to fluctuation of $Q_g$ reads thus:
\beq
	\delta F(t)= {Q_g C'_g \over C^2_g} \delta Q_g(t) \,.
\eeq
In general the variation of the charge on the gate is proportional to the 
variation of the charge on the central island of the SET. 
By an elementary electrostatic calculation (see Appendix \ref{appA})
$\delta Q_g/e=(C_g/C_\Sigma) \delta n$, where $C_\Sigma=C_g+C_L+C_R$ is the sum of the capacitances 
of the central island to all the electrodes and $-n e$ is the total charge on the island. 
In conclusion one finds that 
\beq
	\delta F(t) = F_0 \delta n(t)
\eeq
with 
\beq
	F_0= {Q_g e C'_g \over C_g C_\Sigma}=2{Q_g\over e} {E_C\over L}
\eeq
the force acting on the oscillator when an electron is added to the dot and
with $E_C= e^2/(2C_\Sigma)$ the Coulomb energy of the SET.
Note that $F_0$ is a crucial parameter, since it constitutes the 
electro-mechanical coupling constant.\cite{pistolesi_current_2007,micchi_mechanical_2015}
One can estimate the typical value of $F_0$: $Q_g/e =$ 10$-$100, $E_C= 1$ K, $L=100$ nm, 
thus $F_0 \approx 10^{-11}$-$10^{-12}$ N. 

The correlation function of the stochastic force acting on the resonator 
[$S_{F}(t)= \langle \delta F(t) \delta F(0) \rangle$]
is thus simply proportional to the correlation function of the 
charge on the island [$S_{\delta n}(t)= \langle \delta n(t) \delta n(0) \rangle$]:
\beq
	S_{F}(t)=F_0^2 S_{n}(t)
		\,,
\eeq
that can be calculated by the standard method of the master equation.
For the case of a metallic dot see for instance Refs.~\onlinecite{pistolesi_current_2007,weick_euler_2011}.
Its Fourier transform has a Lorentzian form with a width on the scale of $\Gamma$. 
Thus this force act as a white noise on the slow oscillator.

Let's now turn to the displacement correlation function. 
In order to evaluate it we use a simple Langevin approach.\cite{blanter_single-electron_2004,mozyrsky_intermittent_2006}
We neglect the driving, since we are interested in the low frequency response.
The Langevin equation reads
\beq
	m\ddot x + m \gamma \dot x + kx = \delta F(t),
	\label{langevin}
\eeq
where $m$  is the (effective mass) of the oscillator mode considered, $\gamma$ the damping coefficient, and
$k$ the effective spring constant.
The stochastic force generated by the electrons is also at the origin of the damping 
coefficient. 
In general other effects participate, but close to the degeneracy point of the SET, when the current 
is maximal, the electronic contribution to the damping can dominate, as observed experimentally 
in Ref. \onlinecite{ganzhorn_dynamics_2012}. 
We will assume thus that $\gamma$ is due only to the electronic damping. 
In equilibrium the fluctuation-dissipation theorem gives
\beq
	S_{\delta F}(\omega=0)=2 \gamma m k_B T .
\eeq 
For finite $eV\gg k_B T$, the system is out of equilibrium and 
one has to evaluate explicitly $\gamma$ and $S_{\delta F}$
from a direct calculation of $S_{F}(\omega)$.
As shown in Ref. \onlinecite{clerk_quantum-limited_2004} 
$2 m\hbar  \gamma =\left.dS_{F}(\omega)/d\omega \right|_{\omega=0}$.
One can then always define an effective temperature by the relation 
$S_{F}(\omega=0)=2 \gamma m k_B T_{\rm eff}$,
since the oscillator has a very sharp response in frequency 
and the correlation functions are flat on that scale, one 
can always interpret the ratio of the fluctuation and the dissipation 
as an effective temperature. 
In the case of the SET it has been shown that the typical value of 
$k_B T_{\rm eff}$ is of the order of $eV$.\cite{armour_classical_2004}

The Langevin equation (\ref{langevin}) can then be solved by Fourier transform giving
\beq
	S_x(\omega) =\langle x(\omega) x(-\omega) \rangle=
	{F_0^2 S_{ n} (\omega) \over m ^2 |\omega_m^2 -\omega^2-i\gamma \omega|^2} 
\eeq
and in particular in the low-frequency limit:
\beq
	S_x(\omega=0)={F_0^2 S_{ n} (\omega=0) \over m^2 \omega_m^4}
	\,.
\eeq
We can now use the expansion (\ref{expansion})  to find the lowest order 
contribution of the stochastic fluctuations of $x(t)$ to the current.
We denote these fluctuations $\delta x(t)$ to distinguish them from the time-dependent average 
induced by the external driving:
\beq
	\delta I(t)= {\partial I \over \partial n_g} {V_{g0} C'_g\over e}  \delta x(t) +\dots
	\label{expansionx}
	\,.
\eeq
The back-action current noise is then 
\beq
	S^{\rm ba}_{I}(\omega)= 2 \left({\partial I\over \partial n_g} {F_0 n_g  \over k L } \right)^2
	 S_{n}(\omega=0)
	 \label{SIIx}
\,.
\eeq
As discussed in Refs. \onlinecite{pistolesi_self-consistent_2008,bruggemann_large_2012}
the mechanical back-action noise can be very strong and induce effective giant Fano factors. 

Finally the measurement added noise can be obtained as is done for the 
amplifiers,\cite{clerk_introduction_2010} by dividing the fluctuation 
of the current signal by the amplifier gain squared. 
This gives:
\beq
	S_{x}^{\rm add} = {S_{\rm mx}\over \lambda^2}={S^{\rm shot}_{I}+S^{\rm ba}_{I} \over 4 \lambda^2}
	\label{SxxImp}
		\,.
\eeq
This quantity gives the upper bound on the detection sensibility, since the limitations considered are intrinsic to 
the detection method. 
We will evaluate explicitly these quantities for two specific models in sections \ref{sec5} and 
\ref{sec6}.

%%%%%%%%%%%%%%%%%%%%%%%%%%%%%%%%%%%%%%%%

\section{Fast oscillator}
\label{sec4}

%%%%%%%%%%%%%%%%%%%%%%%%%%%%%%%%%%%%%%%%

In this section we relax the condition  $\omega_m \ll \Gamma $ for 
the calculation of the mixing current. 
We assume $\hbar \Gamma, \hbar \omega_m  \ll k_B T$, the electronic transport is 
then decribled by sequential transport and we will find the mixing current to lowest 
non-vanishing order in the amplitude of the oscillating 
field by making use of a master equation description.

Let's begin by introducing in some details the electron tunnelling description. 
We assume that the only available charge states on the island are those associated 
with two charge states $Ne$ and $(N+1)e$.
We will call these two states 0 and 1.  
The state of the SET is thus fully described by the probabilities 
of one of these two state to be realized: $P_n$, with $n=0,1$. 
We define $\Gamma^{L+(-)}$ as the rate for adding (subtracting) one electron
on (from) the central island  through the left tunnel junction.
Similarly we define $\Gamma^{R+(-)}$ for the right junction.
We define also $\Gamma^\alpha=\Gamma^{L\alpha}+\Gamma^{R\alpha}$, with $\alpha=\pm$,
$\Gamma^L=\Gamma^{L+}+\Gamma^{L-}$, $\Gamma^R=\Gamma^{R+}+\Gamma^{R-}$,
and $\Gamma^T=\Gamma^+ + \Gamma^-$.
The master equation for the the probability reads ($\dot P\equiv dP/dt$):
\beqa
	\dot P_0 &=&-\Gamma^+ P_0+\Gamma^- P_1\\
	\dot P_1&=&\Gamma^+ P_0-\Gamma^- P_1. 
\eeqa
Using the conservation of probability ($P_0+P_1=1$)
we are left with
\beq
	\dot P_0 = -\Gamma^T P_0 + \Gamma^- \label{masterEq}
		\, .
\eeq
We consider now that the rate equations are modulated by two oscillating 
parameters, in our specific case $V$ and $n_g$.
We expand in power series of the amplitude of oscillation the rates keeping 
only the lowest orders:
\beq
	\Gamma^\alpha(t)
=
	\Gamma^{\alpha(0)}(t)
+
	\Gamma^{\alpha(1)}(t)
+	\Gamma^{\alpha(2)}(t)
+ \dots
\eeq
where $\alpha$ stands for any of the previously introduced labels, and the 
term into parenthesis indicates the order in the expansion.
As far as the driving frequency is smaller than the temperature,
$\hbar \omega_i \ll k_B T$,
the explicit expression of the time-dependent rates can be obtained by 
that for the static case by substituting the time-dependent fields:\cite{bruder_charging_1994} 
for instance
$\Gamma^\alpha(t)=\Gamma^\alpha(a(t),b(t))$, where 
$a=a_0+a_1(t)$, $b=b_0+b_1(t)$, and $a_1(t)=a_1\cos(\omega _1 t)$, 
$b_1(t)=b_1 \cos(\omega_2 t)$.
One can then expand to second order in the time dependent part of the two parameters to obtain:
\beqa
	\Gamma^\alpha(t)&=&\Gamma^\alpha
	+
	{\partial\Gamma^\alpha   \over \partial a}  a_1(t)
	+
	{\partial  \Gamma^\alpha\over \partial b}  b_1(t)
	+
	{1\over 2}{\partial^2 \Gamma^\alpha\over \partial a^2}  a_1^2(t) 
		\nonumber \\
	&&
	+	
	{\partial^2 \Gamma^\alpha\over \partial a \partial b} a_1(t) b_1(t) 
	+	
	{1\over 2} {\partial^2 \Gamma^\alpha\over \partial b^2}  b_1^2(t) 
	+\dots  \,. \nonumber
\eeqa
The expansion up to second order can then be rearranged in a Fourier series:
\beqa
	\Gamma^\alpha(t)
	&=&
	\Gamma^\alpha_{00}
	+
	\sum_{n=-1,1}
	\left[ \Gamma_{n,0}^{\alpha(1)} e^{i n \omega_1 t}
	+
	 \Gamma_{0,n}^{\alpha(1)} e^{i n \omega_2 t}
	 \right]
	 \nonumber \\
	 &&
	 +
	\left[\Gamma_{1,-1}^{\alpha(2)} e^{i (\omega_1-\omega_2) t} + {\rm cc} \right]
	+ \dots \label{expGamma}
\eeqa
where the static part $\Gamma^\alpha_{00}$ has contributions of zero and second order 
in the driving fields.
The notation $\Gamma_{n,m}^{\alpha(p)}$ indicates a contribution of order $p$ in the 
driving intensity.
Concerning the time dependent second order terms, we keep only the interesting part at the mixing-current 
frequency $\omega_\Delta$.

We look for a solution of the master equation in terms 
of the stationary Fourier components 
\beq
	P_0(t)=\sum_{n,m} A_{n m} e^{i(n \omega_1 + m \omega_2 )t}
	\label{expP}
	\,.
\eeq
This gives for each Fourier component the equation:
\beq
	(i n \omega_1+i m \omega_2) A_{n m} + \sum_{n',m'} \Gamma^T_{n'm'}
	A_{n-n', m-m'} -\Gamma^-_{nm}
	=0
	\,.
 \eeq
We further expand the $A$ coefficients writing: 
\beq
	A_{n m}=\sum_{p=0}^\infty A^{(p)}_{n m} ,
\eeq
where again $p$ indicates the order in the driving fields.
This leads to a set of equations that can be solved recursively.
The zeroth-order one reads: 
\beq
	(i n \omega_1+i m \omega_2) A^{(0)}_{n m} + \Gamma^{T(0)}_{00}
	A_{n, m}^{(0)} -\Gamma^{-(0)}_{00}\delta_{n,0}\delta_{m,0}=0
	\, .
\eeq
It gives immediately the static solution:
\beq
	A^{(0)}_{n m}=\delta_{n,0}\delta_{m,0} {\Gamma^{-(0)}_{00}\over \Gamma^{T(0)}_{00}}
	\,.
\eeq
For the next two orders we obtain:
\beq
	A^{(1)}_{n m} 
		=
		{ 
		\Gamma^T_{00} \Gamma^{-(1)}_{nm}-\Gamma^{T(1)}_{nm}\Gamma_{00}^{-(0)}
		\over
		\Gamma^{T(0)}_{00}( in\omega_1+i m \omega_2+\Gamma^{T(0)}_{00}) 
		}
	\,,
\eeq
and
\beq
	A^{(2)}_{n m} 
		={\Gamma^{-(2)}_{nm}
		-{\Gamma^{T(2)}_{00}} A^{(0)}_{n,m}
		-\sum_{n'm'}
		{\Gamma^{T(1)}_{n'm'}} A^{(1)}_{n-n',m-m'}
		\over
		 i n \omega_1+i m \omega_2+\Gamma^{T(0)}_{00}
		}
	\,.
\eeq
The non-vanishing terms up to order two are 
$A^{(0)}_{0,0}$,
$A^{(1)}_{\pm1,0}$, $A^{(1)}_{0,\pm1}$, 
$A^{(2)}_{0,0}$, $A^{(2)}_{\pm2,0}$, $A^{(2)}_{\pm 1,\pm 1}$, and $A^{(2)}_{0,\pm2}$.
As usual for the Fourier transform of real functions the following relation holds: 
$A^*_{n,m}=A_{-n,-m}$.

Let us now consider the particle current. 
It can be expressed in terms of $P$ and $\Gamma$, for instance, 
on the left junction (note that this expression does not include the displacement current):
\beq
	I(t)/e=\Gamma^{L+}P_0-\Gamma^{L-} P_1=\Gamma^{L}P_0-\Gamma^{L-} 
	\label{expI}
	\,.
\eeq
Substituting the expansion (\ref{expP}) into \refe{expI} we obtain 
for $I$ a similar expansion to \refe{expP}.
The first three orders read:
\beqa
I_{nm}^{(0)}/e&=& \left[ \Gamma_{00}^{L(0)}A_{00}^{(0)}-\Gamma_{00}^{L-(0)} \right]\delta_{nm}\delta_{n0}
\\
I_{nm}^{(1)}/e&=& \Gamma_{nm}^{L(1)} A_{00}^{(0)}+ \Gamma_{00}^{L(0)} A_{nm}^{(1)}
-\Gamma_{nm}^{L-(1)}
\\
I_{nm}^{(2)}/e &=& 
 \Gamma_{00}^{L(0)} A_{nm}^{(2)}
+\sum_{n' m'}  \Gamma_{n-n',m-m'}^{L(1)} A_{n'm'}^{(1)} 
\nonumber \\
&& 
+\Gamma_{nm}^{L(2)} A_{00}^{(0)}
-\Gamma_{nm}^{L-(2)} .
\label{mixingI2}
\eeqa

The mixing current is given by 
\beq
	I^c_{\rm mx}={\rm Re} I_{1,-1}/2 \quad ,\quad
	I^s_{\rm mx}=-{\rm Im} I_{1,-1}/2
	\,.
\eeq
In order to simplify the expressions obtained above 
we use the fact that in general $\omega_1 \approx \omega_2 \equiv \omega_D$ 
so that even in the fast oscillator limit $|\omega_1-\omega_2| \ll \Gamma_{00}^{T(0)}$. 
This gives the approximate expressions:
\beqa
		A^{(1)}_{10} 
		&=&
		{ 
		\Gamma^T_{00} \Gamma^{-(1)}_{10}-\Gamma^{T(1)}_{10}\Gamma_{00}^{-(0)}
		\over
		\Gamma^{T(0)}_{00}( i \omega_D+\Gamma^{T(0)}_{00}) 
		}
		\\
		A^{(2)}_{1,-1} 
		&=&
		{\Gamma^{-(2)}_{1,-1}
		-\Gamma^{T(1)}_{1,0}  A^{(1)}_{0,-1}
		-\Gamma^{T(1)}_{0,-1}  A^{(1)}_{1,0}
		\over
		 \Gamma^{T(0)}_{00}
		}
\eeqa
One can see that the residual $\omega_D$-dependence is due to the relaxation time of the charge in 
the island. 
As expected it disappears for $\omega_D \ll \Gamma^{T(0)}_{00}$.
The contribution from $I^{(1)}_{1,-1}$ vanishes since 
$\Gamma_{1,-1}^{\alpha(1)}=0$.
The interesting part is the contribution of second order which 
reads:
\beqa
I_{1-1}^{(2)} &=& 
 \Gamma_{00}^{L(0)} A_{1,-1}^{(2)}
+ 
\Gamma_{1,0}^{L(1)} A_{0,-1}^{(1)} 
+
\Gamma_{0,-1}^{L(1)} A_{1,0}^{(1)} 
\nonumber \\
&& 
+\Gamma_{1,-1}^{L(2)} A_{00}^{(0)}
-\Gamma_{1,-1}^{L-(2)}
\, .
\label{mixingI2-bis}
\eeqa
One can verify  that for $\omega \ll \Gamma^{T(0)}_{00}$ expression \refe{mixingI2-bis}
reduces to $\partial I^2/\partial a \partial b$ recovering the 
standard results for the mixing-current [cf. expressions (\ref{MixSin}) and (\ref{MixCos})].

In the opposite limit of $\omega \gg \Gamma^{T(0)}_{00}$ the first order correction to the 
charge variation vanishes ($A^{(1)}_{1,0}\rightarrow 0$): the charge has not 
the time to follow the driving. 
Only a second order correction survives $A^{(2)}_{1,-1}=\Gamma^{-(2)}_{1,-1}/ \Gamma^{T(0)}_{00}$.
The residual time dependence at the mixing frequency is only due to the direct 
modulation of the tunneling rates ($\Gamma_{1,-1}^{\alpha(2)}$).
The final expression for $I_{1,-1}$ in the limit $\omega\rightarrow \infty$ reads:
\beq
{I_{1-1}^{(2)}}_{\rm fast} = 
\Gamma_{00}^{L(0)} 
 {\Gamma^{-(2)}_{1,-1}
		\over
		 \Gamma^{T(0)}_{00} }
+\Gamma_{1,-1}^{L(2)} A_{00}^{(0)}-\Gamma_{1,-1}^{L-(2)}
\label{mixingI2-fast}
\eeq

In the following two sections we consider explicitly the case of a metallic dot and of 
a single electronic level dot and we derive explicit expressions for the 
mixing current, its fluctuation and the response function in the high-frequency regime.

%%%%%%%%%%%%%%%%%%%%%%%%%%%%%%%%%%%%%%%%

\section{The metallic dot single-electron transistor}
\label{sec5}

%%%%%%%%%%%%%%%%%%%%%%%%%%%%%%%%%%%%%%%%

The expression for the tunnelling rate are well known for a 
metallic dot in the Coulomb blockade regime.\cite{ingold_g.-l._charge_1992} 
For convenience of the reader, we report in the appendix a very short 
derivation of the electrostatic relations.
We consider only the two states with $N$ and $N+1$
electrons.

\subsection{Low temperature case}

We begin by discussing the low temperature case $k_B T \ll eV \ll E_C$ 
where $E_C= e^2/2C_\Sigma$ is the Coulomb energy.
In this case there are only two non-vanishing rates (for $V>0$) 
\beqa
	\Gamma_L^+(N)&=&\Gamma_o(v+\tilde n_g) \theta(v+\tilde n_g)
	\label{GammaL}
	\\
	\Gamma_R^-(N+1)&=&\Gamma_o(v-\tilde n_g) \theta(v-\tilde n_g)
	\label{GammaR}
\eeqa
where $\Gamma_o=1/R C_\Sigma$, $v=(C+C_g/2)V/e$ and $\tilde n_g= C_g(x)V_g/e-N-1/2$,
we assume a symmetric device with tunneling resistance $R$.
The stationary solution to the master equation (\ref{masterEq}) and the
stationary current (\ref{expI}) read
\beq
	P_1^{\rm st}= {\tilde n_g+v \over 2v} \, ,
	\quad
	I = e\Gamma_o {v^2-{\tilde n_g}^2 \over 2v} \,,
	\label{P1eq}
\eeq
both equations valid for $|\tilde n_g|<v$. 
The current vanishes continuosly for  $|\tilde n_g|\geq v$ 
while the probability is 1 for $\tilde n_g>v$ and 0 for $\tilde n_g<-v$.

The driving amplitudes in terms of the dimensionless variables introduced read
$v_1$ and $n_{g1}$.
Note that the dependence of the rates on $v$ and $\tilde n_g$ is non-analytic
for $\tilde n_g=\pm v$, this gives a constraint on the amplitude of the 
oscillations since the Taylor expansions are not valid if the parameters cross
this values. This gives the constraints $|\tilde n_g \pm n_{g1}|<v$ and 
$v - v_1> \tilde n_g$, that can be written $n_{g1}, v_1 < v-\tilde n_g$.
Using \refe{GammaL} and \refe{GammaR} we can readily obtain the 
non-vanishing coefficients
of the expansion (\ref{expGamma}):
$\Gamma_{00}^{L+} =\Gamma_o(v+\tilde n_g),$ 
$\Gamma_{10}^{L+}=\Gamma_o e^{i\varphi} n_{g1}/2$, 
$\Gamma_{01}^{L+} =\Gamma_o{v_1}/{2}$, 
$\Gamma_{00}^{R-} =\Gamma_o(v-\tilde n_g)$,
$\Gamma_{10}^{R-}=-\Gamma_oe^{i\varphi} {n_{g1}}/{2}$,
$\Gamma_{01}^{R-} =\Gamma_o{v_1}/{2}$.
For $ \omega_1 \approx \omega_2 = \omega_D$ we obtain a very 
simple expression for the component  $I_{1,-1}$:
\begin{equation} 
	I_{1-1}= e \Gamma_o \frac{\tilde n_g v_1 n_{g1} e^{-i\varphi}} {\tilde \omega_D^2+ 4 v^2}
\end{equation}
here we defined $\tilde \omega_D=\omega_D/\Gamma_o$. 
One finds thus a Lorentzian behaviour, the amplification factor 
decreases quite rapidly for large frequency driving $\omega_D$.
The main reason for the reduction of sensitivity is the 
incapacity of the charge in the dot to follow the driving signal. 
The crossover value for the frequency is $\omega_D \approx V/Re$, 
above this value one cannot use anymore the adiabatic approximation
for the relaxation of the charge on the dot. 
It simply coincides with the frequency for which one electron per driving period
crosses the device.
For instance for $\omega_m=100$ MHz, $R=10^5$ Ohm,  
for voltage below a mV the corrections due to the retardation 
of the charge on the dot becomes relevant
This regime has been observed in the experiment presented in
Ref. \onlinecite{benyamini_real-space_2014}, where the crossover 
from slow to fast oscillator has been investigated by a fine tuning
of the tunnelling resistances.

The amplification factor for the mechanical quadratures is thus:
\begin{equation} 
\lambda = 
{e \Gamma_o \over L}  
\frac{n_g  \tilde n_g v_1}{(\tilde \omega_D^2+ 4 v^2)}
\, .
\end{equation}
It is maximum for $\tilde n_g=\pm v$, but one should also take into 
account the constraint on the amplitude of $v_1<v-|\tilde n_g|$.
One way to take that into account is to set $v_1= v-|\tilde n_g|$, this is 
the maximum allowed value for the driving amplitude, and since the signal 
increases linearly with $v_1$, it gives the maximum value for $\lambda$.
This gives:
\begin{equation} 
\lambda = 
{e \Gamma_o \over L}  
\frac{n_g  \tilde n_g (v-|\tilde n_g|)}{(\tilde \omega_D^2+ 4 v^2)}
\, .
\end{equation}
The maximum of $\lambda$ as a function of the gate voltage is obtained for $\tilde n_g= \pm v/2$ and 
its value (for $\tilde \omega_D \ll v$) is 
\begin{equation}
	\lambda= {e \Gamma_o n_g \over 16 L}  
	\label{lambda}
\eeq
independently of $v$. 
For a typical device one has $n_g\approx 100 $ 
$L\approx 1 \mu $m, $\Gamma_0=10^{11}$ Hz leading to 
$\lambda \sim $ 0.1 A/m. \cite{moser_nanotube_2014,moser_ultrasensitive_2013}

The gain is only a part of the detection, one has  also to evaluate the noise. 
For that we need the two contributions considered in the section \ref{sec3}.
The Fano factor has been obtained in Ref.~\onlinecite{korotkov_intrinsic_1994} (cf. Eq.~41 there):
\beq
	{\cal F}= {{\Gamma_L^+}^2+{\Gamma_R^-}^2 \over (\Gamma_L^+  +\Gamma_R^-)^2}
	=
	{v^2+{\tilde n_g}^2 \over 2 v^2}
\eeq
it varies between 1/2 and 1.
The shot noise becomes thus:
\beq
	S_{II}^{\rm shot}=e^2 \Gamma_o {v^4- {\tilde n_g}^4 \over 2 v^3}
	\label{Sshot}
	\,.
\eeq 

To obtain the contribution of the displacement fluctuation we need to calculate 
the charge noise correlation function: 
$S_n(t)=\langle \delta n(t) \delta n(0) \rangle$. 
This symmetric correlator can be obtained by the conditional probability 
$P(1t |10)$ that the dot it occupied at time $t>0$ with the condition that 
it was occupied at time 0: 
$S_n(t)=[P(1t |10)-P_1^{\rm st}]P_1^{\rm st} $. 
Solving the master equation with the initial condition $P_1=1$ one 
finds 
\beq
	P(1t |10)=1+(P_1^{\rm st}-1)(1-e^{-\Gamma^T t}),
\eeq
with
$P_1^{\rm st}=\Gamma^+/\Gamma^T$.
By Fourier transforming we obtain: 
\beq
	S_n(\omega)=P_1^{\rm st}(1-P_1^{\rm st})  {2 \Gamma^T  \over \omega^2 + {\Gamma^T}^2} 
\eeq
As expected the correlation function is flat for $\omega \ll \Gamma^T$, the required 
low frequency correlator reads then:
\beq
	S_n(\omega=0)=2 \Gamma^+ \Gamma^-/{(\Gamma^T)}^3.
\eeq
In the specific case of low temperature one obtains thus 
$S_n=(v^2-{\tilde n_g}^2)/(4\Gamma_o v^3)$. 

In the typical working regime of a SET $V\ll V_g$, and 
$n_g\approx N$. 
Using the \refe{Qi} one finds that $Q_g/e \approx n_g \approx N $. 
We thus have  $F_0=2N E_C/L$.
Collecting all the terms we can substitute into \refe{SIIx} to obtain:
\beq
	S_{I}^{\rm ba}=2 e^2 \Gamma_o N^4 \left(E_c\over k L^2\right)^2 {(v^2-{\tilde n_g}^2){\tilde n_g}^2 \over v^5}
	.
	\label{Sdisp}
\eeq
The ratio of the mechanical to the shot noise is thus:
\begin{equation} 
	\frac{S_{I}^{\rm ba}}{S^{\rm shot}_{I}}=\left(\frac{E_c}{kL^2}\right)^2 \frac{4N^4 \tilde n_g^2}{v^2(v^2+\tilde{n}_g^2)}
	.
\end{equation}
For large mechanical coupling ($L$ small and $N$ large) the mechanical noise dominate even if 
for small $\tilde n_g$ it is always suppressed, due to the vanishing of $ \partial I/ \partial n_g$.

From Eqs (\ref{lambda}), (\ref{Sshot}), and (\ref{Sdisp}), we obtain the seeked added noise as defined by 
\refe{SxxImp}. 
In order to study its dependence on the different parameters it is convenient to introduce the two 
dimensionless variables 
$\nu\equiv \tilde n_g/v$ and the dimensionless coupling constant
$\delta\equiv(E_C/k L^2) (N^2/v)= \epsilon_P/eV$, where 
$\epsilon_P=F_0^2/k$ is the energy scale of the electromechanical coupling.\cite{armour_classical_2004,doiron_electrical_2006,pistolesi_current_2007}
The added noise then reads:
\beq
	S_{x}^{\rm add}= {E_c \over k \Gamma_0} f(\nu,\delta)
	\, ,
		\label{SxxImp}
\eeq
with
\beq
	f(\nu, \delta)={2 (1+\nu) [4(\delta^2+1) \nu^2+1] \over \nu^2(1-\nu)\delta}
		\,,
\eeq
and $0<\nu<1$. The function diverges for $\nu\rightarrow 1$ due to the 
fact that we have to limit the amplitude of the voltage modulation and diverges for 
$\nu\rightarrow 0$ due to the vanishing of the amplification factor. 
The minimum added noise is thus always for values of $\nu$ between 
0 and 1. 
In the weak coupling limit, for $\delta \ll 1$, one finds that the minimum is at $\nu \approx 0.54$
and reads 
\beq
		S_{x}^{\rm add} \approx 14.8 {E_c/k \over \Gamma_0 } {eV \over \epsilon_P} .
\eeq
For strong coupling, $\delta\gg1$, instead the minimum is close to $\nu=0$ with a value for 
\beq
	S_{x}^{\rm add} \approx 8 {E_c \over k\Gamma_0} \left({\epsilon_P\over eV}\right)^2.
\eeq
In both cases the noise diverges when $\delta$ becomes very small or very large. 
In the weak coupling limit the added noise is dominated by the 
current noise (imprecision noise), in the strong coupling it is instead given essentially by the 
back-action noise. 
As usual\cite{clerk_introduction_2010} the optimal situation is in the middle for $\delta\approx 1$. 

\begin{figure}
\centering
\includegraphics  [angle=0, width=0.5\textwidth]{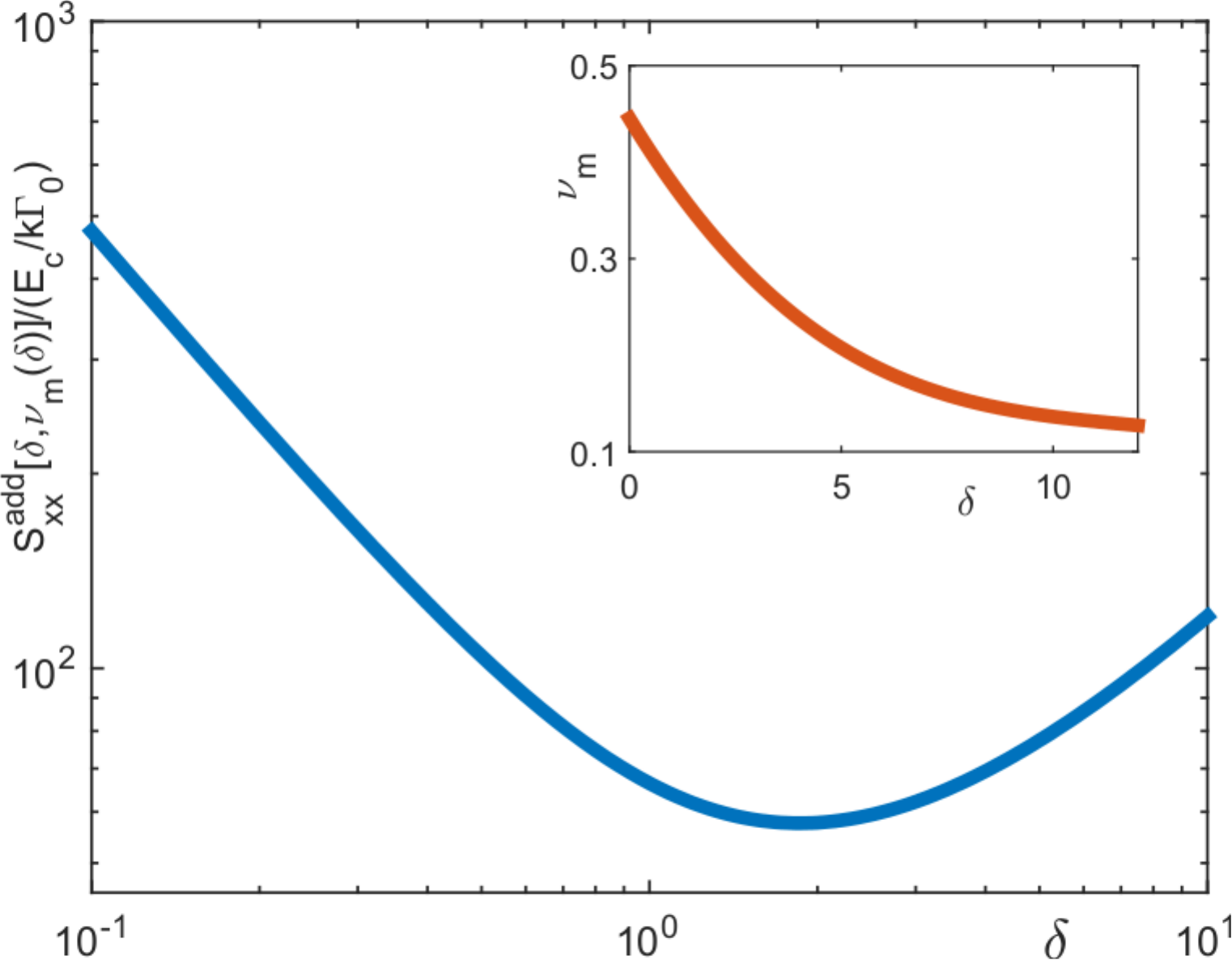}
\caption{$S_{xx}^{\rm add}$ as a function of $\delta=\epsilon_P/eV$ for 
$\nu(\delta)$ minimizing the function . In a inset the value of $\nu$ that minimizes the function
for given $\delta$ [$ \nu_m(\delta)$].}
\label{figMinimum}
\end{figure}

In \reff{figMinimum} we plot $S_{x}^{\rm add}[\delta, \nu_{\rm m}(\delta)]$, where 
$\nu_{\rm m}(\delta)$ is the value of $\nu$ that minimizes $S_{xx}^{\rm add}$
for given $\delta$. 
We thus find that the absolute minimum for the added noise is obtained for 
$\nu=0.32$ and 
$\delta=1.857$ and reads 
\beq
	S_{xx}^{\rm add}= 57.61 {E_c \over k \Gamma_0}
\eeq
This is the ultimate sensitivity that can be obtained with this device in ideal conditions,
when all other sources of imprecisions have been eliminated. 
 Inserting typical values of $E_c\approx 10 K$, $\Gamma_0\approx 10^{11} Hz$, $k=10^{-5} N/m$ one 
 obtains the value of $S_{xx}^{\rm add} \approx 10^{-26} m^2/Hz$.
 One should regard this value with some caution. 
 Let's consider the value of the coupling that is required to obtain this sensitivity.
 The optimal value of $\delta$ is for $eV \approx \epsilon_P$.
 As discussed in the litterature (see for instance Ref. \onlinecite{pistolesi_current_2007}, where 
 this energy is called $E_E$) this scale  determines the value at which the system
 undergoes a current blockade. 
It is difficult to reach this limit (since one needs also $k_B T \ll \epsilon_P$) in metallic SETs.
On the other side $\epsilon_P$ of the order of 0.3 K has been observed in suspended carbon 
nanotubes.\cite{benyamini_real-space_2014}
 The dramatic effects expected at low tempereture on the mechanical resonators have been discussed 
 recently.\cite{micchi_mechanical_2015,micchi_electromechanical_2016}
 This extreme limit need to be reconsidered, since the resonating frequency of the 
 resonator is renormalized by the coupling, and the added noise induced by the oscillator 
 is expected to be more effective. 
 In particular the oscillator becomes strongly non-linear close to the transition.

 We can estimate in a simple way the effect of the softening of the mechanical resonator following 
 Ref.~\onlinecite{pistolesi_current_2007}. 
 The correction to the variation of the energy reads\cite{blanter_single-electron_2004} 
 \beq
 	\Delta E^{\pm} \rightarrow \Delta E^{\pm} \pm F_0 x
\eeq
this changes the form of $P^{\rm st}_1$ given by \refe{P1eq} as follows:
\beq
	P_1^{\rm st}= {\tilde n_g+xF_0/(2E_C)+v \over 2v}
	\,.
\eeq
 Substiting into the equation for the average force $F_0 P^1_{\rm st} $ and
 taking the derivative with respect to $x$ one obtains the renormalized 
 spring constant:
 \beq
 	k'=k(1-\delta)
 	\,.
 \eeq
The instability appears for $\delta=1$, where two new stable solutions bifurcate.
The only change in our previous calculations is the value of $k$ entering \refe{SxxImp}:
 \beq
	S_{xx}^{\rm add}= {E_c \over k \Gamma_0} {1\over 1- \delta} f[\nu,\delta/(1-\delta)]
	\,.
\eeq
 Repeting the minimization procedure we find that the minimum is now for $\delta=0.48$ 
 holding the value of $132.7 (E_c/k \Gamma_0)$. 
 Thus the renormalization of the resonating frequency reduces the precision of a 
 factor of 2, leaving space for high sensitivity detection.

 The actual limitation in current experiments will be the value of the coupling, since 
 in practice the typical temperature reached in experiments on metallic quantum dot 
 is much larger than $\epsilon_P$.
In the following section  we considerthe detection at finite temperature and low voltage.

\subsection{Finite temperature case} 
\label{subsecFiniteTemp}

 Let us now consider the finite temperature case $eV \ll k_B T \ll E_C$.
 In this case we have to take into account the four possible 
 tunnelling processes that change the charge on the dot from the $N$ to the 
 $N+1$ state (cfr. Appendix). 
 The respective rates read:
 \beqa
 	\Gamma_{L(R)}^+(N)&=&\Gamma_{\rm Th} h[(\mp eV-2 \tilde n_g E_C)/k_B T], \\
 	\Gamma_{L(R)}^+(N+1)&=& \Gamma_{\rm Th} h[(\pm eV+2 \tilde n_g E_C)/k_B T],
\eeqa	
with $h(y)=-y [1-e^{y}]$ and $\Gamma_{\rm Th}=k_BT/e^2R$.
We consider the low bias voltage limit $eV/k_BT \ll 1$.
In this limit the expression for the current \refe{expI} becomes:
 \beq
 	I= e\Gamma_{\rm Th} 
 	 {eV\over 2 k_B T}  g(2 \tilde n_g E_C/k_B T) ,
\eeq
where 
\beq
	g(y)={h_+h'_-+h_+'h_- \over  h_++h_- } = {e^y y \over e^{2y}-1}
\eeq	 
and $h_\pm=h(\pm y)$.
From the expression of the current we obtain 
\beq
	{\partial^2  I \over \partial V \partial \tilde n_g}=-{E_C \over R k_B T} g'(2 \tilde n_g E_C/k_B T) ,
\eeq
(for brevity, we omit in the following the arguments of $g$ and of the other functions of $y=2 \tilde n_g E_C/k_B T$)
with the amplification factor:
\beq
	\lambda= { n_{g0} \Gamma_0 eV_1 \over 4 L k_B T  } g'
	\approx {F_0 V_1 \over 8 R k_B T} g'
	\label{lambdaT}
	.
\eeq
The factor $g'(x)$ has a maximum for $y=1.16$ for which 
it holds the approximate value $0.154$. 
Thus tuning $\tilde n_g = 0.58 k_B T /E_C$ allows to obtain the maximum 
value of the amplification factor. 
Comparing this value to \refe{lambda}, valid for $k_B T \ll eV$, we see that 
the amplification factor is reduced by the term $eV_1/k_B T \ll 1$.

The shot noise at low frequency reads:\cite{korotkov_intrinsic_1994}
  \beq
  	S_{I}^{\rm shot}
  	=
  	e^2
  	\left[ 
  	{\Gamma_L^+\Gamma_R^- + \Gamma_L^- \Gamma_R^+   \over \Gamma_T}
  	-2 {(\Gamma_L^+ \Gamma_R^- - \Gamma_L^- \Gamma_R^+)^2 \over \Gamma_T^3}
  	\right]
 \,.
\eeq
For small $V$ the first term (thermal noise) dominates and gives:
\beq
	S_{I}^{\rm shot}=e^2\Gamma_{\rm Th}  {h_+ h_- \over h_+ + h_-}
	\label{SIIshot}
	\,.
\eeq
The charge noise in the same limit reads
 \beq
 	S_{n}={1\over \Gamma_{\rm Th}} {h_+ h_-\over (h_+ + h_-)^3}
 	\,.
 \eeq
 
 From the expression of the back-action noise (\ref{SIIx}) we see that for 
 $V\rightarrow0$ there is (apparently) no back action of the measurement. 
 It is possible to set $V=0$ and exploit its modulation around 0 to detect the motion of the oscillator. 
 But in this case we need to consider the next order in the expansion (\ref{expansion}).
 For $V=0$ we have:
 \beq
 	\delta I={\partial I \over \partial n_g \partial V} \delta n_g V_1+\dots \, .
\eeq
From this we have for the current-current correlator:
\beq
	\langle \delta I(t_1) \delta I(t_2) \rangle = \left({\partial G\over \partial n_g}\right)^2 V_1(t_1) V_1(t_2) 
	\langle \delta n_g(t_1) \delta n_g(t_2) \rangle
	\,,
 \eeq
 where $G=dI/dV$ for $V=0$ is the conductance.
The product of the two $V_1$ terms gives an oscillating term 
depending on $t_1+t_2$ that averages to zero and a second term proportional to $\cos[\omega_2(t_1-t_2)]$.
Using $\delta n_g(t)= (C'_g V_g/e) \delta x(t)$ we have
\beq
	S_I^{\rm ba}= {1\over 2} \left({\partial G\over \partial n_g} \right)^2 (V_{g0} C_g' V_1/e)^2 S_x(\omega_2)
	\label{BackActionN}
	\,.
\eeq
Typically $\omega_2\approx \omega_m$, we thus assume that it is resonant in order to evaluate the case of 
maximal back-action:
\beq
	S_I^{\rm ba}=  %{{g'}^2 h_+h_-\over 8(h_++h_-)^3} {1\over \Gamma_{\rm Th}} \left( {\epsilon_P Q V_1 \over k_B T R}\right)^2
	{{g'}^2 h_+h_-\over 16(h_++h_-)^3} e^2 \Gamma_0 {\epsilon_P^2  Q^2  (e V_1)^2 \over (k_B T)^3 E_C } 	  
	,
	\label{SIIba}
\eeq
with $h^{\rm ba}={(g')}^2 h_+h_-/(h_++h_-)^3$ and $Q=\omega_m/\gamma$ the oscillator quality factor.

Adding the two sources of current noise \refe{SIIba} and \refe{SIIshot} we obtain for the added noise:
\beq
	S_x^{\rm add} = {E_C \over k \Gamma_0}
	  \left[ \alpha^{\rm ba} {\epsilon_P Q^2 \over k_BT} + \alpha^{\rm shot} {(k_B T)^3\over \epsilon_P (eV_1)^2}\right]
	  \,,
\eeq
with the numerical factors
$\alpha^{\rm ba} =4 h_+ h_-/(h_++h_-)^3$ and $\alpha^{\rm shot}=32 h_+ h_-/[(g')^2(h_++h_-)]$.
Choosing the value $\tilde n_g=1.60$ that maximizes $\lambda$ their values are
$\alpha^{\rm ba} = 0.23 $ and $\alpha^{\rm shot}=449$.

The minimum of the added noise is obtained for 
\beq
	\epsilon_P =\left({\alpha^{\rm shot} \over \alpha^{\rm ba}}\right)^{1/2}  {(k_BT)^2 \over Q eV_1}
	\label{minEP}
		\,,
\eeq
with a minimum noise of 
\beq
	S_x^{add}=2  {E_C \over k \Gamma_0} \left(\alpha^{\rm ba} \alpha^{\rm shot}\right)^{1/2}  Q {k_B T \over eV_1}
	\,.
\eeq
Since $eV_1/ k_B T \ll1$, at best we can set this ratio to 0.1. 
This gives for the optimal value of the coupling 
\beq
	{\epsilon_P \over k_BT } \approx   {441 \over Q}
\eeq
and the minimum of the added noise 
\beq
	S_x^{\rm add}= 203  {Q E_C \over k \Gamma_0 }
	\,.
\eeq

Some comments are at order. 
First we assumed that the frequency driving the voltage bias is resonant with the oscillator. 
This is un upper limit to the back action, in particular if $Q\gg 1$ this condition is not fulfilled and the back action 
will be reduced. 
For the non-resonant case it is sufficient to use the above results with $Q\approx \omega_m/\omega_{\Delta}$, reducing enormously the minimum
added noise, to the expenses of finding a much larger coupling constant. 
The second comment concern the value of the coupling constant $\epsilon_P$ necessary to reach the minimum.
One can see that even with the assumption of resonant back action it is relatively large.
For a typical $Q\approx 10^4$ one finds $\epsilon_P/k_BT \approx 0.04$.
To our knowledge the largest value of the ratio $k_B T/ \epsilon_P$ is $\approx 0.017$ has been reported 
in Ref. \onlinecite{benyamini_real-space_2014}.
Since as soon as $Q\gg 1$ it is possible to avoid resonant back-action, 
in most cases the main limitation is to reach large values of $\epsilon_P$.

%
%
% \begin{figure}
%\centering
%\includegraphics  [angle=0, width=0.5\textwidth]{FigMinT2}
%\caption{finite T minimum prefactor as a function of delta when y is at the minimum and (yellow) without minimization by fixing %$y=2.16$. }
%\label{figMinimumT}
%\end{figure}

It is interesting to compare the shot-noise contribution of the added noise with the resonant brownian motion fluctuations:
\beq
	S_x^{\rm B}(\omega_m)=2 {k_B T\over k \gamma}
	\,.
\eeq
The ratio reads:
\beq
	{S_x^{\rm add} \over S_x^{\rm B}} = { \alpha^{\rm shot} \over 2} {E_C \gamma \over \epsilon_P \Gamma_0} 
	\left( {k_B T\over eV_1}\right)^2
	\,.
\eeq
Detection of brownian motion can then be done for 
$\epsilon_P/E_C> 2 \times 10^{4}  \gamma/\Gamma_0$ (where we assumed as before $eV_1/k_B T=0.1$). 
For instance in Ref.~\onlinecite{moser_ultrasensitive_2013} $\gamma/\Gamma_0 \approx 10^{-8}$ 
% Gamma_0 (there called W in out) 7 10^10 Hz, gamma=omega0/Q omega0=2pi 5 MHz Q=5 10^4 
% 
%	gamma/Gamma_0 = 10^-8 
%
allowing the detection of the brownian motion fluctuations even for very weak coupling. 
For a rough estimate of the coupling in that experiment one can use the expression given in Ref.~\onlinecite{micchi_mechanical_2015}
$\epsilon_P/k_B T\approx 2 \delta \omega_m/\omega_m$, where $\delta \omega_m$ is the 
modulation of the resonating frequency near the degeneracy point (see Fig.~3 in Ref.~\onlinecite{moser_ultrasensitive_2013}). 
For Ref.~\onlinecite{moser_ultrasensitive_2013} one finds $\epsilon_P \approx 16$m K to be compared to $E_C$ of the order of 
10K. 
Notwithstanding the low value of the coupling constant, the resolution is largely sufficient to detect the Brownian motion of the 
carbon nanotube.

\section{The single-electronic level SET}
\label{sec6}

When the temperature and the voltage bias is much smaller than the electronic level separation the 
rates for electron transfer reads:\cite{beenakker_theory_1991}
\beqa
  \Gamma_{L(R)}^+ &=& \Gamma_{L(R)0} f_F[(\epsilon-\mu_{L(R)})/k_BT],
  \\
  \Gamma_{L(R)}^-&=&\Gamma_{L(R)0} \left[1-f_F[(\epsilon-\mu_{L(R)})/k_BT ]\right] \,,
 \eeqa
where $f_F(y)=1/(1+e^{y})$ if the Fermi function, 
$\epsilon$ is the level position,  $\mu_L(R)$ is the left (right) chemical potential, and 
$\Gamma_{L(R)0}$ are the transfer rates.
For simplicity in the following we choose  $ \Gamma_{L0}=\Gamma_{R0}=\Gamma_{0}$.
The modulation of the gate voltage leads to the time-dependence 
$\epsilon(t)=\epsilon_{0}+\epsilon_1(t)$ 
of the electronic level energy $\epsilon$ with 
\beqa
  \epsilon_{0} &=& \epsilon_{d0}+e C_gV_{g0}/C_{\Sigma} \,,
  	\\
  \epsilon_1(t) &=& e[C_g'  V_{g0} x(t)+C_g V_{g1}(t)]/C_{\Sigma} \,,
 \eeqa
and  $\epsilon_{d0}$ the position of the electronic level for vanishing $V_g$.
We assume symmetric bias so that the chemical potential read:
\beq
	\mu_{L(R)}(t)=\mu_{L(R)0}+(-)e (V+V_1\cos \omega_2 t)/2
		\,.
\eeq
Following the steps of the previous section we can calculate the current 
\beq
	I={e \Gamma_0\over 2} \left[f_F[(\epsilon-\mu_L)/k_BT]-f_F[(\epsilon-\mu_R)/k_BT]\right]
\eeq
from which we obtain for vanishing $V$ the amplification factor:
\beq
	\lambda= {e n_{g0} \Gamma_0 \over 4 L} {eV_1 E_C \over (k_B T)^2} f''_F(y)  ,
\eeq
where the argument of the Fermi function  is  $y=(\epsilon_0-\mu)/k_BT$, and will be omitted in the following.
The maximum of $f''_F$ is obtained for $y=1.31$ with a value of 0.096.
The thermal part of the shot noise and the charge noise read:
\beqa
 	S_I^{\rm shot} &=& e^2 \Gamma_0 f_F(1-f_F) , \\
 	S_n &=&  {f_F(1-f_F) \over \Gamma_0 } .
\eeqa
Using \refe{BackActionN} for the back-action noise we obtain
\beq
	S_I^{\rm ba} = {f_F(1-f_F) {f_F''}^2 \over 8} e^2 \Gamma_0 
		\left( 
		{eV_1 Q \epsilon_P \over (k_B T)^2}
		\right)^2
		\,.
\eeq
The added noise has thus the form:
\beq
	S_x^{\rm add} = {k_B T\over k \Gamma_0}
		\left[\alpha^{\rm ba} Q^2 { \epsilon_P \over k_B T}
			+ \alpha^{\rm shot} \left({k_B T\over eV_1}\right)^2 { k_B T\over \epsilon_P}
		\right]
\eeq
with $\alpha^{\rm ba}= 2f_F(1-f_F) $ and 
$\alpha^{\rm shot}= 16 f_F(1-f_F)/{f_F''}^2$. 
Their values for $y=1.31$ are $\alpha^{\rm ba}=0.34$ and 
$\alpha^{\rm shot}= 289.2$. 
We find the same value of $\epsilon_P$ for the minimum of the added noise 
in the metallic case [cf. \refe{minEP}], but the minimum of the noise has a different 
expression:
\beq
	 S_x^{\rm add} = 2\left(\alpha^{\rm ba} \alpha^{\rm shot}\right)^{1/2} {(k_B T)^2 \over k \Gamma_0 e V_1}
	 .
\eeq
Essentially the energy scale of the Coulomb blockade is substituted by the temperature, in principle reducing the added noise. 
The conclusion is that the single-level  SET should allow a better resolution of the metallic SET
by a factor $E_C/k_B T$.

\section{Conclusions}

In this work we  have studyed  theoretically the sensitivity of the mixing-current technique. 
We first found general expressions valid when the oscillator resonating frequency is comparable or 
larger of the transfer rate of electrons. 
We find that a reduction of the amplification factor of the order of $(\Gamma_0/\omega_D)^2$ is expected.
This effect should be relatively small in most practical experimental realizations.
We then analysed the fundamental limitations due to the intrinsic noise present in the (current) signal and 
the effect of the back-action fluctuations. 
On general grounds one finds that an optimal value of the electromechanical coupling ($\epsilon_P$) 
exists that minimizes the added noise. 
This value is larger than what is realized in the present experiments, showing that increasing the coupling 
allows to reach higher sensitivity. 
At finite temperature the relevant parameter is the ratio $\epsilon_P/k_B T$ and values of the order of 1 are 
needed to reach the optimal minimum added noise. 
At vanishing temperature the relevant parameter is instead $\epsilon_P/eV$.
In all cases the scale of the sensitivity is given by $E_C/\Gamma_0k$.
Optical means can detect CNTs displacement with good accuracy, even if the small size of the object does not allows to reach the spectacular sensitivity obtained with macroscopic mirrors. 
A sensitivity of $5\cdot 10^{-22}$ m$^2$/Hz as been reported\cite{stapfner_cavity-enhanced_2013} by cavity-enhanced optical detection of CNTs.

We considered only classical fluctuations. It seems difficult to use the mixing technique 
to reach the quantum limit of detection, since the effective temperature of the oscillator, even at vanishing temperature,
is of the order of $eV$ that typically needs to be larger than $\hbar \omega_m$.
On the other side it may be instructive to compare the sensitivity found at vanishing temperature with the 
zero point fluctuations spectrum at resonance: $S^{\rm SQL}_x= 2 \hbar \omega_m/k\gamma$. 
One sees that the ratio to the typical mixing-current technique added noise at zero temperature is
$10^{-2} (\hbar \omega_m/E_C) (\Gamma/\gamma)(\epsilon_P/E_C)$, since 
$\Gamma/\gamma\gg1$, for sufficiently large $\epsilon_P$ the added noise can be of the same order of the 
zero-point fluctuations. 

We conclude that the sensitivity of the mixing technique can still be improved by increasing the electromechanical coupling
till reaching $\epsilon_P$ of the order of the temperature or the Coulomb blockade energy where the back-action will be of the same order of the intrinsic current noise of the device.

\section*{Acknoledgements} 

Y.W. thanks the China Scholarship Council for financial support. 
We thank R. Avriller for useful discussions.

%%%%%%%%%%%%%%%%%%%%%%%%%%%%%%%%%%%%%%%%

\appendix
\section{}

\label{appA}

%
%
%
%%%%%%%%%%%%%%%%%%%%%%%%%%%
%                                                                                %
%                              FIGURE A1                                 %
%                                                                                %
%%%%%%%%%%%%%%%%%%%%%%%%%%%
%
\begin{figure}[tbp] % Requires \usepackage{graphicx}
\includegraphics[width=.85 \linewidth]{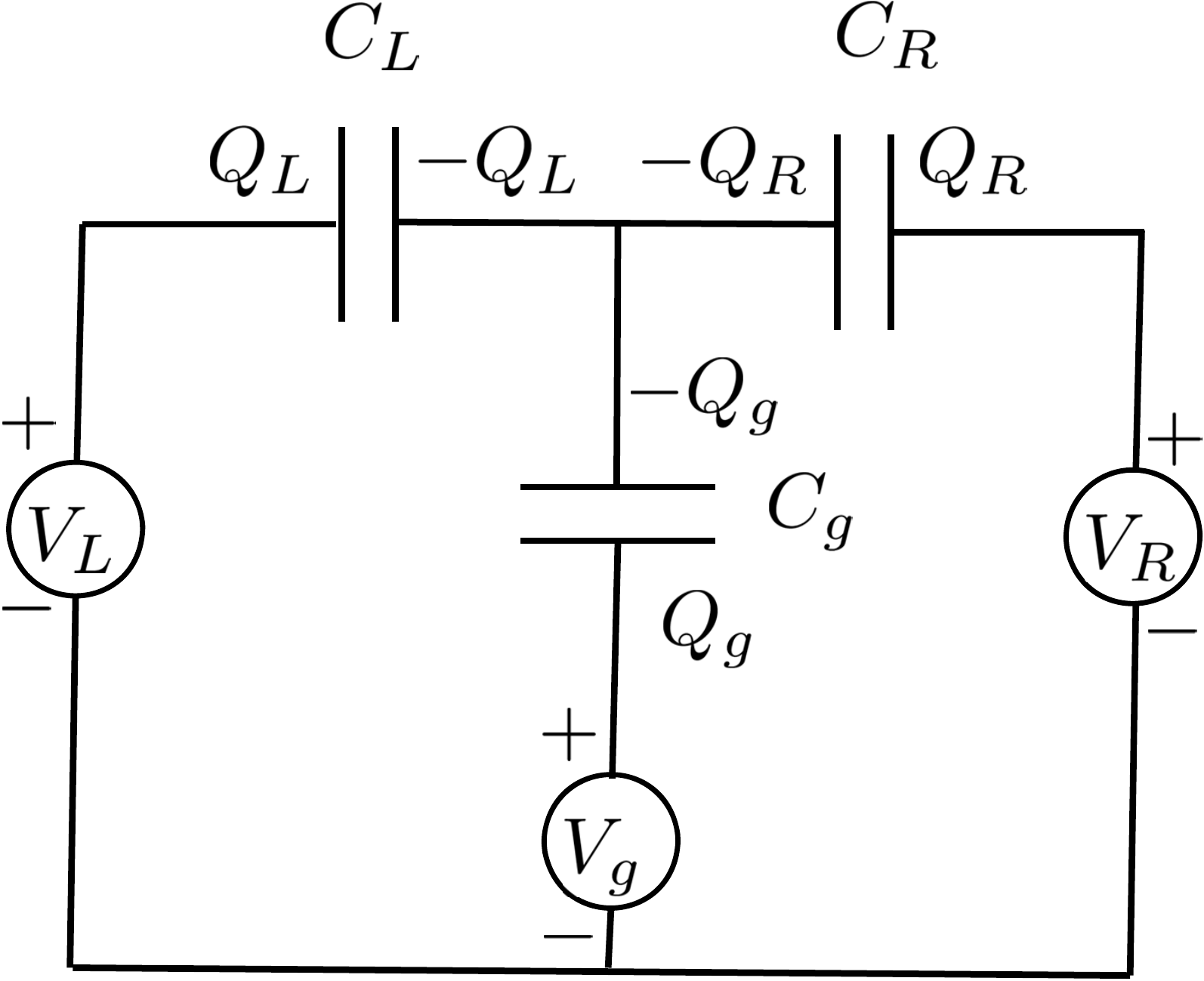}
\caption{Electric scheme of a single electron transistor}
%
%\label{figA1}
\end{figure}
%
%%%%%%%%%%%%%%%%%%%%%%%%%%%%
%

In this appendix we present, mainly for clarifying 
the notation, a brief derivation of the electrostatic 
energy variation for the tunnelling of an electron in a 
single electron transistor.
\cite{ingold_g.-l._charge_1992}
The electric scheme is presented in Fig.3 where 
the potentials of the left, right, and gate leads are 
defined as $V_L$, $V_R$, and $V_g$, respectively.
In the same way the charge on each capacitance (on the leads side) 
is indicated with $Q_i$ with $i=L$, $R$, and $g$.
Defining $V_I$ the potential of the island one has 
\beq
	Q_i=(V_i-V_I)C_i
		\label{Qi}
	\,.
\eeq
Summing the three equations one obtains immediately the expression
for the potential on the island: 
\beq
	V_I=\left(\sum_i C_i V_i+Q\right)/C_\Sigma,
	\label{VI}
\eeq
where $C_\Sigma=\sum_i C_i$ and $Q=-\sum_i Q_i$ is 
the total charge on the island.
The total electrostatic energy $E_e(Q)=\sum_i Q_i^2/2C_i=Q^2/2C_\Sigma+{\rm constant}$,
where the constant term does not depend on $Q$.
From \refe{Qi} and \refe{VI} one then finds that adding a charge 
$q$ on the island will change the charge on each capacitor plate 
of 
\beq
	\delta Q_i=-{q C_i\over C_\Sigma} 
		\,.
\eeq
The total  electrostatic energy variation (final energy minus initial energy) 
for the transfer of an electron 
from the left electrode on the island is then given by the variation of 
the total electrostatic energy plus the the work done by the voltage sources ($-\sum_i V_i \delta Q_i$, 
with $\delta Q_i=e C_i/C_\Sigma$ for $i\neq L$  and $ \delta Q_L = e C_L/C_\Sigma -e$):
\beq
	\Delta E^{+}_L = E_e(Q-e)-E_e(Q)-e\sum_i V_i {C_i \over C_\Sigma}+ e V_L
	\,.
\eeq
The general expression reads then:
\beq
	\Delta E^{\pm}_{L,R} = -e{(-e \pm 2 Q) \over 2 C_\Sigma}  \mp {e\over C_\Sigma}
	 (\sum_i V_i C_i -C_\Sigma V_{L,R})
	\,.
\eeq
The variation of the energy depends only on the difference of the three potentials, we can thus 
choose to express it in terms of $V=V_R-V_L$ and $V'_g=V_g-(V_L+V_R)/2$.
For simplicity we write the expressions in the symmetric case of $C_L=C_R=C$:
\begin{eqnarray}
\Delta E_L^{\pm}&=&\frac{e}{2C_{\Sigma}}(e{\mp}2Q){\mp}\frac{e}{C_{\Sigma}}(C' V+C_g V'_g)\\
\Delta E_R^{\pm}&=&\frac{e}{2C_{\Sigma}}(e{\mp}2Q){\mp}\frac{e}{C_{\Sigma}}(-C'V+C_g V'_g)
\end{eqnarray}
with $C'=C+C_g/2$. Typically $V$ is very small, while $V'_g$ can be very large, in particular 
$V'_g C_g /e=n_g$ is normally regarded as finite, while $C_g\rightarrow 0$ and $V_g'\rightarrow \infty$.
For this reasons we can normally neglect the displacement dependence induced by 
$C_g(x)$ in $C'$ or $C_\Sigma$, while it is necessary to keep the $x$ dependence in 
$C_g(x)$ that appears in the expression $C_g V_g'$.

Let now focus on the four energy variations associated with the change of the number 
of electrons in the dot between the two states $N$  and $N+1$. 
We need $\Delta E_{L,R}^{+}(N)=-\Delta E_{L,R}^{-}(N+1)$ that can 
be explicitly written as:
\beq
	\Delta E^{+}(N)_{L,R}=-{e^2\over C_\Sigma}(n_g\pm v-N-1/2) ,
\eeq
with $n_g=C_g V_g'/e$ and $v=C'V/e$.
The expression of the tunneling rate is obtained then by the Fermi golden rule:
\beq
	\Gamma^{\alpha \pm} = {k_B T \over e^2 R_\alpha}h(\Delta_\alpha^{\pm}/k_B T) 
\eeq
with 	$h(x)=-x/(1-e^{x})$.
In particular for $T\rightarrow 0$ 
the expression for the rate 
becomes simply 
$\Gamma^{\alpha \pm} = -{\Delta_\alpha^{\pm} / e^2 R_\alpha} \theta(-\Delta_\alpha^{\pm})$.
%$k_B T \ll e^2/C_\Sigma
These expressions allow to obtain the tunneling rates necessary for the calculations 
presented in the main text of the paper.

\bibliography{Full}

\end{document}